      [1999/12/01 v1.4c Il Nuovo Cimento]
\documentclass{cimento}

\newcommand{\eiso}{E$_{\rm{iso}}$}
\newcommand{\liso}{L$_{\rm{iso}}$}
\newcommand{\epeak}{E$_{\rm peak}$}
\newcommand{\erad}{E$_{\rm{rad}}$}

\usepackage{graphicx}  
\title{Redshift indicators for gamma-ray bursts}
\author{J-L. Atteia\from{ins:x}}
\instlist{\inst{ins:x} Laboratoire d'Astrophysique, Obs.
Midi-Pyr\'{e}n\'{e}es, 14 Av. E. Belin, 31400 Toulouse, France}
\PACSes{\PACSit{98.70}{Gamma-ray bursts}\PACSit{98.62}{Redshift}}
\begin{document}

\maketitle

\begin{abstract}
The measure of the distances and luminosities of gamma-ray bursts (GRBs) led
to the discovery that many GRB properties are 
strongly correlated with their intrinsic luminosity, leading to the
construction of reliable luminosity indicators.
These GRB luminosity indicators have quickly found applications, 
like the construction of 'pseudo-redshifts', or the measure 
of luminosity distances,
which can be computed independently of the measure of the redshift.
In this contribution I discuss various issues
connected with the construction of luminosity-redshift
indicators for gamma-ray bursts.
\end{abstract}

\section{Introduction}
\label{intro}

Measuring the cosmological redshifts of gamma-ray bursts of all types
(GRBs, XRFs, short-hard bursts) is crucial for our understanding of these 
events. The measure of redshifts tells us the energy output of the source, 
it allows the measure of the burst intrinsic parameters (duration, peak 
energy etc...), and determines the position of the burster in the history
of the universe. Unfortunately, the obtention of spectroscopic redshifts 
requires a succession of non trivial observing steps: reliable localization 
of the prompt emission in X-rays 
or gamma-rays, quick distribution of the alert to the ground, identification
of the afterglow at optical, radio or X-ray wavelenths, and measure
of the redshift of the host galaxy in absorption (when the afterglow
is bright) or in emission (when the afterglow has faded).
This complex sequence of events explains why in February 2005, 
8 years after the discovery of the first afterglow, the web page of 
Jochen Greiner\footnote{http://www.mpe.mpg.de/~jcg/grbgen.html} contains
only 40 spectroscopic redshifts among 263 localized GRBs (of which about 
100 have an optical, x-ray or radio afterglow). The statistics is similar 
for HETE-2 GRBs \cite{vand05}, which has localized 75 events, from
which 25 afterglows have been found, and 14 redshifts measured. 

The fact that we are currently measuring the redshift of one out of six 
localized GRBs implies strong biases in the observed redshift distribution.
For instance GRBs at high redshift are very difficult to detect in the optical 
range, and the lack of GRBs with z $>$ 5 in the current sample may be the consequence of
selection effects. The intrinsic faintness of XRFs makes
them difficult to detect beyond z=1, again biasing their observed 
redshift distribution.
It is precisely this small fraction of GRBs with a redshift and the
likely existence of biases in the observed redshift distribution which
constitute the rationale for the construction of 'redshift indicators'.
In order to complement spectroscopic redshifts usefully, 'redshift
indicators' must provide redshift estimates for most GRBs detected 
at high energies\footnote{In the following, we call 'pseudo-redshifts' these 
redshift estimates.}. 
The construction of such indicators is considered in section \ref{building}.
Section \ref{issues} addresses some open issues in the construction of
redshift indicators.
Possible applications of these indicators are presented in Section \ref{appli}.

\section{Building redshift indicators}
\label{building}

\subsection{Searching standard candles}
\label{candles}

Gamma-ray bursts have a broad dispersion in peak luminosity (by a factor
10$^4$ at least), which prevents using them as standard candles
straightforwardly. After the the measure of a dozen GRB redshifts,
much work has been devoted to the search of luminosity estimators 
with an intrinsic dispersion smaller than the peak luminosity. 
This work brought two complementary results: 

\begin{itemize}

\item \erad, the energy radiated by GRBs when the beaming factor 
of the emission is taken into account\footnote{The 
beaming factor of the emission is usually derived 
from the time at which the light curve of the afterglow shows
a break \cite{rhoa97}. The measure of the fluence of a GRB and of its
redshift can be used to derive \eiso , the isotropic-equivalent
radiated energy. The identification of a jet-break allows
the determination of the opening angle of the jet, giving the beaming factor
and the energy radiated, \erad .}, appears to be significantly
less scattered than the peak luminosity (L$_{\rm iso}$) 
or the energy radiated (\eiso ) measured 
assuming an isotropic emission~\cite{frai01,bloo03}.
The standard deviation of the logarithm of \erad\ is 0.35 dex~\cite{bloo03},
providing a measure of luminosity which approaches a standard candle.

\item Many temporal and spectral GRB properties
are strongly correlated with the radiated energy (either \eiso\ or \erad ).
These correlations permit to use these temporal or spectral properties 
as luminosity indicators, and as redshift indicators by assuming a 
cosmology\footnote{In the following the terms 
luminosity indicators and redshift indicators are used indifferently.
The reader should nevertheless keep in mind that redshift indicators are 
constructed from luminosity indicators, assuming a cosmology.}.
Even if the theoretical justification of these correlations is not 
fully understood yet, they are currently providing interesting
luminosity indicators with typical standard deviations of the order 
of 0.1 dex~\cite{ghir04b, atte04}. 
Aditional details on these luminosity indicators are given in section
\ref{lumi} below.

\end{itemize}

The construction of reliable luminosity indicators has started only recently, 
with the measure of the first GRB redshifts, and significant improvements 
are to be expected in the coming years. 
In order to facilitate the comparison of existing
and future luminosity indicators, we suggest the use a common 
measure of their quality.
We propose to use $\sigma_{\rm DL}$, the dispersion
of log(D$_L^{\rm pred.}$/D$_L^{\rm meas.}$), where
D$_L^{\rm pred.}$ is the luminosity distance predicted by the luminosity
indicator, and D$_L^{\rm meas.}$ is the luminosity distance which is measured.
Using this measure, the pseudo-redshifts proposed by Atteia~\cite{atte03}
have $\sigma_{\rm DL} = 0.15$ dex, for 24 GRBs with a redshift, 
and $\sigma_{\rm DL} = 0.11$ dex, for the subsample of 13 HETE GRBs.

\subsection{Practical luminosity-redshift indicators}
\label{lumi}

From a general point of view, good redshift indicators require
combinations of GRB parameters which have a small intrinsic scatter
and, as possible, a strong dependence on redshift.
Concerning light curves, the two most publicized correlations are the 
lag-luminosity correlation~\cite{norr00},
and the variability-luminosity correlation~\cite{reic01}.
The construction of luminosity indicators based on these relations 
requires light curves with a large number of counts, 
which can only be provided by instruments having large effective area.
Luminosity indicators based on GRB spectra demand the measure of the
three parameters of the Band function usually adopted to fit GRB 
spectra~\cite{band93}. They are based on the so-called \epeak -\eiso\ 
relation~\cite{amat02,atte03,lamb05}, on the \epeak -\liso\ 
relation~\cite{yone04}, or on the \epeak -\erad\ 
relation~\cite{ghir04b}\footnote{Here \epeak\ designates the energy
of the maximum of the $\nu F\nu$ spectrum of the prompt emission.
\eiso\ and \erad\ have been defined in the footnote in section \ref{candles}.}.
These indicators require instruments having a 
broad spectral range (at least two decades), and, in the case of 
the \epeak -\erad\ relation, a good sampling of the afterglow light curve.
These studies have shown that useful luminosity indicators can be constructed
for GRBs, even if they depend on the details of each GRB detector 
(effective area, spectral range, time and energy resolution), 
and must be adapted for each mission.
Recently, the consistency of redshift indicators obtained 
from the light curves and from the spectra has been demonstrated~\cite{ghir05,pizz05}.
Despite the rapid progress of the last years, some open issues remain, which
are presented in the next section.

\section{Open issues}
\label{issues}

\subsection{Validity of luminosity indicators}
\label{validity}
The importance of the empirical relations used in the definition of
luminosity indicators, and the fact that they are based
on a small number of GRBs with redshifts, raised questions on
their overall validity and on their applicability to the entire GRB population.
Various authors attempted to check the validity of the \epeak -\eiso ,
and of the \epeak -\erad\ relations by verifying if, for every GRB 
in the BATSE sample, one can find a redshift which puts it along
the correlations observed for the GRBs with a redshift.
The conclusions of these efforts are diverse: while Lloyd et al. \cite{lloy00}
actually {\it predicted} the \epeak -\eiso\ relation, Nakar \& Piran~\cite{naka05}, 
and Band \& Preece~\cite{band05} found that the majority of BATSE GRBs cannot
follow this relation, but this conclusion was later contradicted by Ghirlanda 
et al.~\cite{ghir05}, and Bosnjak et al.~\cite{bosn05}. 
These contradictory results may reflect 
the difficulty of properly accounting for the uncertainties
in the measure of individual \epeak\ for all GRBs, or the existence
of a class of outliers (see below), treated differently in these studies.

\subsection{Calibration}
\label{calibration}

Luminosity indicators are not well calibrated, essentially due to the small
number of GRBs with a spectroscopic redshift, and to the restricted range of
redshifts measured (z=0.1 to 4.5). The number of GRBs available for
calibration is further reduced by the fact that many GRBs have their
light curves, spectra or afterglows known with insufficient accuracy to
compute luminosity indicators reliably.
With these restrictions in mind, the accuracy of present-day luminosity indicators
is not too bad since the best of them have a standard deviation smaller than 0.1 dex.

\subsection{Outliers}
\label{outliers}

\begin{figure}
\label{durete_fluence}
\center
\includegraphics[width=9.5cm]{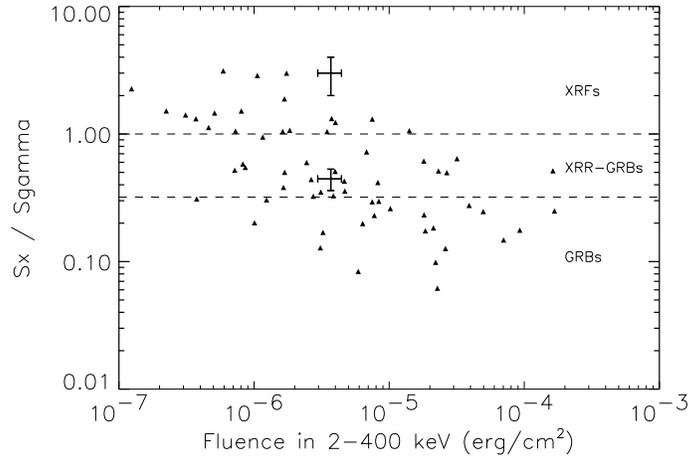}
\caption{Position of GRB 031203 (crosses) in a fluence-softness plot, 
among 63 GRBs detected and localized by HETE-2 \cite{barr04}.
The low (resp. high) softness point reflects the numbers given by 
\cite{sazo04} (resp. \cite{wats04}).
This figure shows that sensitivity considerations cannot be invoked 
to resolve the contradiction raised by GRB 031203: the 
only GRB with a redshift detected with INTEGRAL is an outlier to 
the \epeak -\eiso\ relation, while the 14 GRBs with a redshift 
detected with HETE follow this relation.
}
\end{figure}

The association of GRB 980425 with the energetic supernova 1998bw at z=0.0085
revealed a sub-energetic burst, whose radiated energy is one thousand
times smaller than other GRBs with a redshift. 
As a consequence this burst doesn't fit the 
empirical relations used to define luminosity indicators. 
The nature of the link between GRB 980425 and the bulk of the classical 
GRB population has not been elucidated yet, and we should wait the
detection of additional events of this type to clarify the nature 
of the outliers.
Recently, GRB 031203 has also been considered 
has a potential outlier to the \epeak - \eiso\ and to 
the lag-luminosity relations \cite{sazo04}.
The case of this burst is, however, much less clear than the case
of GRB 980425 because its \epeak\ has not been measured.
Fig. \ref{durete_fluence} illustrates the contradiction raised by
this burst: From the point of view of its prompt gamma-ray emission, 
GRB 031203 appears fully comparable with the GRBs detected with HETE-2;
On the other hand, this burst, the only INTEGRAL GRB
with a redshift, does not follow the \epeak - \eiso\ relation, 
while all the 14 HETE GRBs with a redshift follow this relation.
In our opinion, the situation of GRB 031203 with respect to the empirical 
\epeak - \eiso , and lag-luminosity relations remains unclear.

\subsection{Are redshift indicators applicable to X-ray flashes ?}
\label{xrf}
X-Ray Flashes (XRFs) are soft GRBs with low \epeak , typically
\epeak $< 50$ keV. If they follow the  \epeak -\eiso\ relation, they must
be intrinsically faint and can only be seen at low redshifts.
The redshifts of two XRFs have been measured to date: XRF 020903, at z=0.25, and 
XRF 040701, at z=0.21. These measures put these two bursts on the 
\epeak -\eiso\ relation, and suggest that luminosity indicators for GRBs
could also be applicable to XRFs. The situation is however more complex than this
simple picture since GRBs at high redshift (z=5-10) will also
appear as X-ray flashes.
Due to the absence of GRBs with a measured redshift greater than 5, it is not
clear if luminosity indicators can make the difference between XRFs from
nearby intrinsically soft GRBs and classical hard GRBs at high redshift.
Another puzzle is connected with the lack of supernova light in several 
XRFs~\cite{leva05,sode05}. 
Since all nearby GRBs seem to show SN light~\cite{zeh04},
this observation can be interpreted as an evidence that these XRFs 
lie at redshifts significantly larger
than the values predicted by the redshift indicators 
(but other interpretations are also possible~\cite{leva05}).
Until we know more about the distances of XRFs, it is difficult to
affirm with certainty that GRB redshift indicators are also valid for XRFs.

\subsection{Short-hard GRBs}
\label{short}
The question of redshift indicators for short-hard GRBs is not adressed here
since we have not yet measured the distance of a short-hard GRB.
Furthermore, the recent observation of a very bright gamma-ray
flare from the galactic magnetar SGR 1806-20, suggests that a significant
fraction of the short-hard bursts could be due to extragalactic magnetars~\cite{hurl05}. 
This could make short-hard GRBs fundamentally different from long GRBs
for which luminosity indicators have been shown to work.

\section{Applications}
\label{appli}

We briefly present below some (not all) of the studies 
which have used luminosity-redshift indicators in the last years.

\subsection{Quick redshift evaluation}
\label{highz}
The availability of redshift indicators exclusively based on the prompt emission 
opens the possibility to quickly distribute an evaluation of the distance 
of newly detected GRBs.
This is especially important for GRBs at high redshifts (z$>$5) which 
can only be identified with a rapid and specific follow-up~\cite{atte05}.
In order to facilitate the follow-up strategy of ground observers, the HETE
team has set up a program which performs the spectral analysis of GRBs 
automatically, as soon as the data become available (usually in the minutes 
following the burst), and evaluates
a redshift indicator based on the \epeak - \eiso\ relation.

\subsection{Inferring the star formation rate}
\label{sfr}
GRBs offer an unabsorbed view on the history of the Star Formation Rate (SFR).
The distribution of the pseudo-redshifts of hundreds of BATSE GRBs has
been computed by various authors (see \cite{lloy02,mura03,yone04} and ref. therein).
These studies show that GRBs offer an interesting alternative 
to constrain the SFR beyond z $\sim 3 - 4$, but
we should keep in mind that redshift indicators have
not yet been calibrated in this range of redshifts.

\subsection{Testing the cosmology}
\label{cosmology}
The availability of accurate luminosity indicators for GRBs can be used to
measure luminosity distances independent of the redshift, and to
derive the cosmological parameters. Following \cite{scha03,ghir04a}, various authors
assessed the constraints that GRBs impose on the cosmological parameters.
Whether or not current data are
sufficient in number and accuracy to constrain the cosmological
parameters is the subject of an on-going debate~\cite{dai04,ghir04a,frie05}, 
but there is a strong consensus on the fact that GRBs will become 
a valuable tool in this field in the coming years.

\section{Conclusions and perspectives}
\label{conclusion}
Luminosity and redshift indicators for GRBs will certainly play a
increasing role in the future. The growing number of redshifts
is expected to bring several improvements to this field: a more accurate calibration
of existing luminosity indicators, the construction of new indicators, 
a better understanding of systematic effects (e.g. luminosity evolution) and
of the issues discussed in section \ref{issues}.
The progress in this field may be fast in the SWIFT era thanks to
the small error boxes provided by the XRT, which permit efficient
afterglow searches, and raise the hope that a significant fraction 
of SWIFT GRBs will have spectroscopic redshifts.
A recent good news is that the luminosity indicator
based on the variability of the light curve, 
which was developped for BATSE bursts~\cite{reic01}, 
seems to be directly applicable to SWIFT GRBs~\cite{dona05}.

We would like to conclude with strong incentive to observers to quickly
follow-up at X-ray and NIR wavelengths, GRBs with high pseudo-redshifts. 
The identification of GRBs at high redshifts (z $> 5$) is a difficult
task which requires deep, prompt observations with powerful instruments.
This cannot be done for just any GRB, and the luminosity indicators 
discussed in this paper could be a decisive tool to ensure the success 
of future searches for high-redshift GRBs.


\end{document}